\documentclass[showpacs,aps,pra,amsmath,amssymb,twocolumn]{revtex4}
\usepackage{graphicx}
\usepackage{dcolumn}
\usepackage{bm,color}
\usepackage{txfonts}
\usepackage{amsmath,epsfig}
\usepackage{epstopdf}
\newcommand{\eq}{Eq.~}
\newcommand{\eqs}{Eqs.~}
\newcommand{\fig}{Fig.~}

\newcommand{\cf} {cf.~}
\newcommand{\ug} {\!=\!}

\newcommand{\piu} {\!+\!}
\newcommand{\meno} {\!-\!}

\newcommand{\eg} {e.g.~}

\newcommand{\rref} {Ref.~}
\newcommand{\rrefs} {Refs.~}

\begin{document}

\author{Francesco Ciccarello}
\affiliation{NEST, Istituto Nanoscienze-CNR and Dipartimento di Fisica e Chimica, Universit$\grave{a}$  degli Studi di Palermo, via Archirafi 36, I-90123 Palermo, Italy}
\pacs{03.65.Yz, 42.50.Lc}

\title{Waveguide-QED-based measurement of a reservoir spectral density}

\date{\today}
\begin{abstract}

The spectral density (SD) function has a central role in the study of open quantum systems (OQSs). We discover a method allowing for a ``static" measurement of the SD -- i.e., it requires neither the OQS to be initially excited nor its time evolution tracked in time -- which is not limited to the weak-coupling regime. This is achieved through one-dimensional photon scattering for a zero-temperature reservoir coupled to a two-level OQS via the rotating wave approximation. We find that the SD profile is a universal simple function of the photon's reflectance and transmittance. As such, it can be straightforwardly inferred from photon's reflection and transmission spectra. 
\end{abstract}

\maketitle

\section{Introduction}

Mostly because of progress in quantum technologies that is easing access to a variety of {\it single} small systems \cite{haroche}, the interest in open quantum systems (OQSs) \cite{petruccione,cohen,huelga}, the study of which became topical about thirty years ago, has further strengthened over the last few years. 

A key concept in the study of OQSs is the {\it spectral density} (SD) function \cite{petruccione,cohen,huelga,breuer,lambro}. At any given frequency, the SD essentially measures the interaction strength between the OQS and a  reservoir mode at that frequency weighted by the corresponding density of states of the reservoir. Currently, the concept of SD is growing in importance also due to the increasing interest in non-Markovian (NM) OQS dynamics (such as exciton transport in protein complexes \cite{shabani} where dedicated numerical methods are used to compute the SD \cite{aspuru}). Indeed, structured (namely non-flat) SDs in general entail that the celebrated Kossakowski--Lindblad Markovian master equation (KLME) is not effective \cite{petruccione,cohen,huelga,breuer}. This is a linear first-order differential equation having the system's state as the only unknown and depending on a set of rates (\eg the familiar spontaneous emission rate of an atom). The open dynamics governed by the KLME is the prototype of a quantum Markovian, namely ``memoryless", dynamics. Despite the easiness to handle it, the KLME arises from a number of approximations. As such, it can be quite ineffective in a number of relevant, known scenarios featuring non-negligible NM effects \cite{petruccione}. In such cases, only the knowledge of the full reservoir {\it spectral density} (SD) \cite{petruccione,cohen,huelga,breuer,lambro} guarantees a reliable description of the OQS dynamics. This relies on the crucial property that if the SD is known then the OQS dynamics is fully determined. As a major consequence, the knowledge of the SD is key to designing strategies to hamper decoherence in quantum information processing \cite{nc}. 

Measuring the SD thereby is a task of utmost importance. A possible method \cite{dissi} is to measure the relaxation rate of a probe qubit -- i.e., a two-level system embodying the OQS -- as a function of its Bohr frequency (when this is tunable). This approach relies on the Fermi golden rule (FGR), hence on the assumption that the QOS-reservoir interaction is weak (weak-coupling regime). Other schemes \cite{dephase}, which address purely dephasing noise, exploit external pulsing to modify the OQS dynamical evolution.  Measurements on the OQS are then used to infer the underlying SD. Such methods are dynamical in nature in that the diagnostic process underpinning them is essentially the OQS time evolution. This typically brings about, in particular, the need for initialising and measuring the OQS in suitable states.

Can one devise a SD measurement strategy with no need for triggering a dynamical evolution of the probe OQS (``{\it static"} measurement) and which is effective beyond the weak-coupling regime? In this Letter, we discover that this is achievable for an important class of quantum reservoirs. This encompasses dissipative, zero-temperature baths coupled to a two-level OQS via the rotating wave approximation (RWA). Prominent environmental models extensively investigated in the literature are included, \eg lossy cavities and photonic band-gap mediums coupled to a quantum emitter \cite{petruccione,lambro}. The method for such static SD measurement is spectroscopic in nature: it exploits light scattering from the OQS, the outcomes of which are used to extract informations on the SD associated with the OQS dressed with its own reservoir. As a distinctive trait of the scheme is that it employs light that is constrained to travel in a one-dimensional (1D) waveguide. We find that reflection and transmission spectra, which can be recorded through standard intensity measurements, are enough for fully reconstructing the SD in a surprisingly straightforward fashion. This conclusion relies on an equation that directly maps the SD into a simple combination of reflectance and transmittance of the probing photon. The OQS needs not be initially excited neither its dynamics tracked in time, which embodies the {static} nature of the SD measurement method.

Photon scattering from quantum emitters (even a single one) in 1D waveguides, which we harness as the scheme diagnostic tool, is currently a hot field of research, often dubbed {\it waveguide Quantum ElectroDynamics} (QED) \cite{SF1,SF-PRA,switch,lukin,sore}. 
Technologic advancements make such processes by now experimentally observable, or next to be so, in a broad variety of different setups, such as open transmission lines coupled to superconducting qubits \cite{wallraf,tsai,delsing-NJP} or nanowires (alternatively, photonic-crystal waveguides) coupled to quantum dots \cite{claudon}  (for a more comprehensive review of possible implementations see \eg\rrefs\cite{sore,review}).
The 1D confinement of light gives rise to unique interference effects such as the perfect reflection of a single photon from a quantum emitter \cite{SF1}.
There is growing evidence that the rich physics of waveguide QED can be harnessed for a number of promising applications in photonics, \eg light switches \cite{switch} and single-photon transistors \cite{lukin}, as well as QIP, such as quantum gates \cite{zheng,ciccarello}. The scheme to be presented here further witnesses the potential of waveguide QED.

\section{Environmental model}

 We consider a two-level OQS called $S$ (e.g.~an artificial atom) in dissipative contact with a quantum reservoir $R$ with the joint Hamiltonian modelled as ($\hbar\ug1$ throughout) \cite{petruccione, lambro,cohen}
\begin{eqnarray}
\hat H_{S\!R}\ug\omega_0\,\hat\sigma_+\hat \sigma_-+\!\sum_i \omega_i\,\hat b^\dag_i\hat b_i+\sum_i \mu_i \left(\hat b_i \hat\sigma_+\!+\! \hat b_i^\dag \hat \sigma_-\right)\,.\label{SR}
\end{eqnarray}
\begin{figure}
\begin{center}
\includegraphics[width=0.6\linewidth]{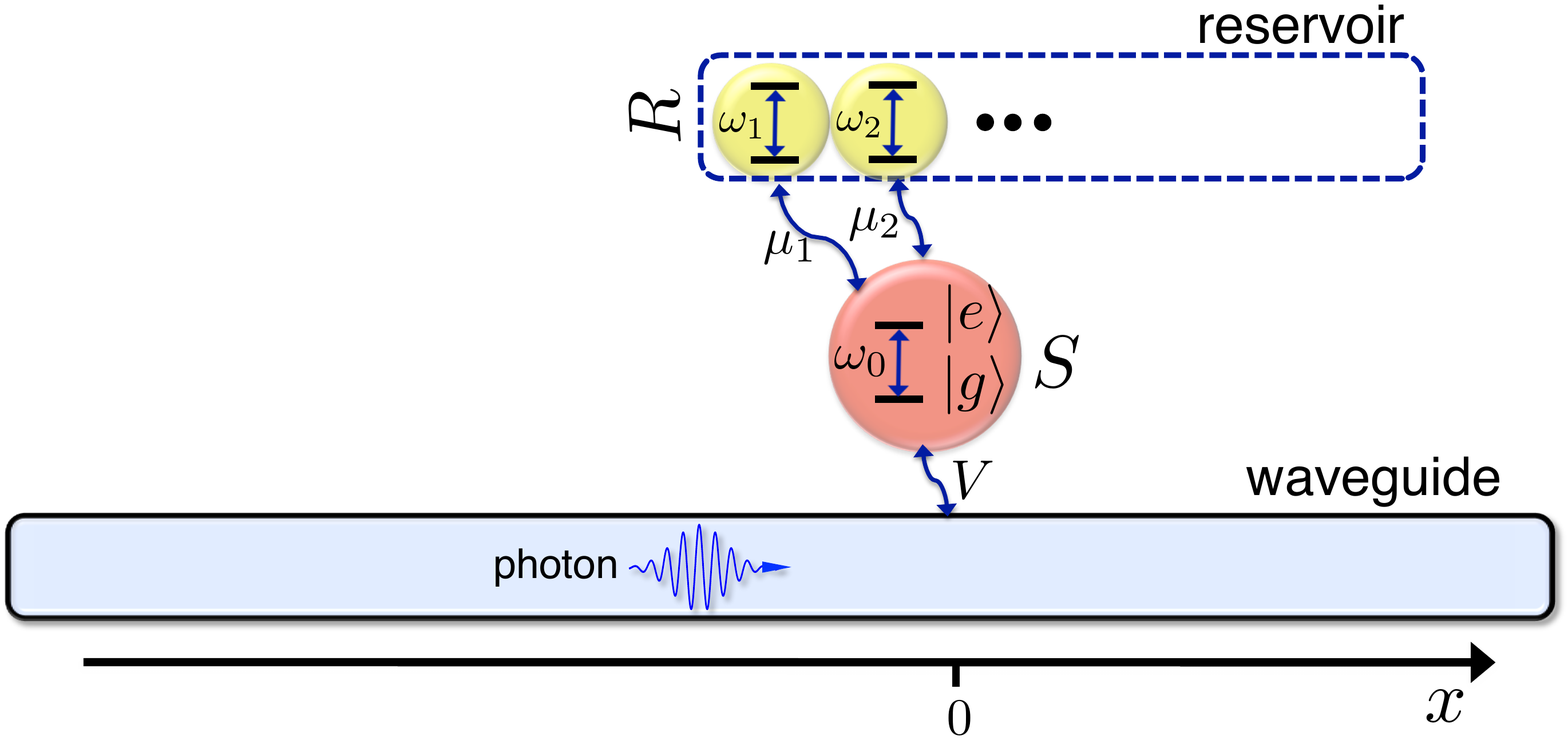}\end{center}\vspace{-.3cm}
\caption{(Color online) Setup for the SD measurement.}
\end{figure} 
Here, $\omega_0$ is the energy separation between the $S$'s excited and ground states $|e\rangle$ and $|g\rangle$, respectively, while $\hat \sigma_-\ug\hat \sigma_+^\dag\ug|g\rangle\langle e|$ are the usual ladder spin operators. $R$ instead comprises a very large numbers of independent modes, the $i$th of which is a harmonic oscillator with associated frequency $\omega_i$ and bosonic annihilation and creation operators $\hat b_i$ and $\hat b_i^\dag$, respectively. In the last sum of \eq(\ref{SR}), the $i$th term accounts for the interaction (under RWA) between $S$ and the $i$th mode of $R$ with corresponding coupling strength $\mu_i$. A sketch of $S$ and $R$ is shown in \fig 1 (top part). Under the usual assumption that $R$ features a continuum of modes instead of a discrete set, $\omega_i$ becomes the continuous frequency $\omega$. Accordingly, $\mu_i\!\rightarrow\!\mu(\omega)$ and $\sum_i\!\rightarrow\!\int \!d\omega \,\rho(\omega)$, where $\rho(\omega)$ is the reservoir's density of states.

Consider now the emission process where $S$ is initially excited while $R$ is in the vacuum state $|{\rm vac}\rangle_R$ (i.e., at zero temperature) and call $\varepsilon(t)$ the probability amplitude to find $S$ still in the excited state $|e\rangle$ at time $t$, i.e., $\varepsilon(t)\ug_S\!\langle e|_R\!\langle {\rm{vac}}|\Psi(t)\rangle_{S\!R}$ with $|\Psi(t)\rangle_{S\!R}$ the joint state of $S$--$R$.
An equivalent representation of $\varepsilon(t)$ is its Laplace transform (LT) $\tilde\varepsilon(z)\ug\int \!dt \,\varepsilon(t)e^{izt}$, where $z$ is a complex variable. 
It can be shown \cite{lambro,cohen} that the general solution for $\tilde\varepsilon(z)$ reads 
\begin{eqnarray}
\tilde\varepsilon(z)=\frac{1}{z-\omega_0\meno\int \!d\omega' \frac{J(\omega')}{z-\omega'}}\,,\label{gee}
\end{eqnarray}
where $J(\omega)$ is the SD defined as $J(\omega)= \rho(\omega)\left[\mu(\omega)\right]^2$.
\eq(\ref{gee}) shows that $\tilde\varepsilon(z)$, hence $\varepsilon(t)$, is ``shaped" by $J(\omega)$: as anticipated, the SD function fully determines the $S$ open dynamics. Note that, at the lowest-order, \eq(\ref{gee}) yields $|\varepsilon(t)|^2\!\simeq\!e^{-2\pi J(\omega_0)t}$ and hence the decay rate \cite{nota-FGR}
\begin{equation}
\gamma_{eg}\ug 2\pi J(\omega_0)\,,\label{FGR}
\end{equation}
i.e., the well-known FGR mentioned in the Introduction. As anticipated in the Introduction, using \eq(\ref{FGR}) to measure $J(\omega_0)$ via $\gamma_{eg}$ is ineffective whenever perturbation theory breaks down. A paradigmatic instance illustrating such drawback is an atom emitting into a photonic band gap (PBG) material for $\omega_0$ lying within a reservoir band gap but close to a band edge \cite{lambro,kurizki}. In this case, $J(\omega_0)\ug0$. 
Using \eq(\ref{FGR}) one would instead find $J(\omega_0)\!\neq\!0$ since it is known that, although incomplete, a significant atomic decay anyway occurs \cite{lambro,kurizki}.
In contrast, the method to be presented here is effective even in such conditions.

\section{Photon scattering from $S$}

 To acquire information on $S$ ``dressed" by its own reservoir $R$, we send a photon on $S$ and study the resulting scattering process (see \fig1). As it is constrained to travel through a 1D waveguide, the photon can be either strictly scattered off -- reflected or transmitted -- or irreversibly absorbed by the $S$-$R$ joint system. Later on, we will see that the possibility of such absorption is key to the scheme working principle. To describe the scattering process, we add further terms to Hamiltonian (\ref{SR}) to include the waveguide field $F$ \cite{SF1}. The total Hamiltonian of $S$, $R$ and $F$ reads
\begin{equation}
\hat{H}\!=\!\hat{H}_{S\!R}+\!\!\int \!\!{\rm d}k\,\upsilon |k| \,\hat a^\dag\!(k)\,\hat a(k)+ V \!\!\!\int \!\!{\rm d}x \,\delta(x)\,[\hat{c}(x)\,\hat\sigma_+\!+\!{\rm H.c.}]\,.\label{H}
\end{equation}
Here, the first integral is the free field Hamiltonian, where $\hat{a}(k)$ [$\hat{a}(k)^\dag$] is a field bosonic operator annihilating (creating) a photon of wave vector $k$. We have assumed that the waveguide features a linear dispersion law, namely the photon's energy $\omega$ depends on its wave vector $k$ along the waveguide axis $x$ as $\omega\ug \upsilon |k|$ with $\upsilon$ being the light group velocity. The second integral in (\ref{H}) instead accounts for the $S$-$F$ coupling occurring at the $S$ position $x\ug0$ (see \fig1) and features real-space field operators $\hat c(x)$ and $\hat c^\dag(x)$, where $\hat{c}(x)$ [$\hat{c}^\dag(x)$] annihilates (creates) a photon at position $x$. Hence, the last integral in (\ref{H}) means that a photon lying at $x\ug0$ can be absorbed by $S$ with the latter being promoted to state $|e\rangle$ (or the inverse process). The coupling strength associated with such process is measured by parameter $V$. Note that, unlike $R$ which is the (unknown) reservoir to be probed, $F$ can be regarded as the ``probing" reservoir. The latter is fully Markovian with a flat SD given by $(V^2/\upsilon)/\pi$ [this is due to the presence of $\delta(x)$ in the second integral of \eq(\ref{H}), reflecting a uniform coupling to the waveguide modes, and the assumed dispersion law linearity].

A photon with wave vector $k\!>\!0$ is sent towards $S$ when the initial state of $S$--$R$ is $|g\rangle_S|\rm{vac}\rangle_R$ (subscripts $S$ and $R$ will be omitted henceforth). Note that unlike the emission process corresponding to \eq(\ref{gee}) here both $S$ and $R$ are initially {\it unexcited}. As usual in quantum scattering problems, we now search for a stationary state of the joint system $S$--$R$--$F$
and enforce that the corresponding energy eigenvalue be $\omega\ug\upsilon k$, namely the same as the incoming photon energy. As only a single photon is sent and $\hat{H}$ does not feature counter-rotating terms, the state to seek lies in the single-excitation sector of the total Hilbert space. Thus
\begin{eqnarray}
\!|\Psi\rangle\ug\!\sum_{\eta=\pm}\!\!\int\!\!dx\,\psi_\eta(x)\,\hat{c}^\dagger_\eta\!(x)|{ {\rm vac}}\rangle|g\rangle\piu \!\sum_i \beta_i\,\hat b_i^\dag |{ {\rm vac}}\rangle|g\rangle\piu \alpha\,|{\rm vac}\rangle |e\rangle\,,\label{ansatz}
\end{eqnarray}
where $|\rm{vac}\rangle\ug|\rm{vac}\rangle_R|\rm{vac}\rangle_F$ is the state where both $R$ and $F$ are in the respective vacuum states. Here, $\hat{c}_+(x)$ [$\hat{c}_-(x)$] annihilates a right- (left-) moving photon at $x$, the associated creation operator being $\hat{c}_+^\dag(x)$ [$\hat{c}_-^\dag(x)$], while $\psi_+(x)\ug [\theta(-x)\piu t\theta(x)]e^{ikx}$ and $\psi_-(x)\ug r\theta(-x)e^{-ikx}$ reflect the usual scattering ansatz \cite{SF1} with $r\!$ ($t$) the photon's reflection (transmission) coefficient. 
Imposing now, as anticipated, that $\hat H|\Psi\rangle\ug\omega|\Psi\rangle$ yields through standard methods \cite{SF-PRA} the set of coupled equations for $\psi_\pm(x)$, $\{\beta_i\}$ and $\alpha$ 
\begin{eqnarray}
\mp i\upsilon\,\frac{ {\rm d}\psi_{\pm}}{{\rm d}x}(x)+V\delta( x)\alpha&=&\omega \,\psi_{\pm}(x)\,,\label{psip}\\
\omega_0\alpha\piu V\left[\psi_{+}(0)\piu\psi_{-}(0)\right]\piu\sum_i\mu_i\,\beta_i&=&\omega \,\alpha\,,\label{alpha}\\
\omega_i\,\beta_i\piu \mu_i\,\alpha&=&\omega\,\beta_i\,.\label{betai}
\end{eqnarray}
Note that in \rref\cite{SF-PRA} the authors focused on the implications of these equations on photon transport by restricting to the special case of a reservoir $R$ with a {\it flat} SD (Markovian reservoir). Here, instead, we tackle the measurement problem of the SD. Thereby, the SD is left fully {\it unspecified} throughout.

We will next eliminate $\{\beta_i\}$ and $\alpha$ so as to end up with a closed equation in $\psi(x)\ug\psi_+(x)\piu\psi_-(x)$ (a similar task was carried out in \rref\cite{ciccarello} but without the reservoir $R$). Subtracting \eq(\ref{psip}) for $\psi_+(x)$ from the analogous equation for $\psi_-(x)$ yields
$i\upsilon\partial_x{\psi}(x)=\omega  \left[\psi_-(x)\meno\psi_+(x)\right]$.
Upon further derivation in $x$, we get $i\upsilon\partial_{x}^{2}\psi(x)\ug\omega  \left[\partial_x\psi_-(x)\meno\partial_x\psi_+(x)\right]$. Replacing next $\partial_x\psi_\pm(x)$ as given by \eq(\ref{psip}), this becomes
\begin{eqnarray}
\frac{ {\rm d}^2\psi}{{\rm d}x^2}(x)+k^2\,\psi(x)\ug\frac{2k V}{\upsilon}\alpha \,\delta(x)\,,\label{psip2}
\end{eqnarray}
where we used $\omega\ug\upsilon k$.
\eq(\ref{psip2}) alongside \eqs(\ref{alpha}) and (\ref{betai}) now form a set of equations in the unknowns $\{\psi(x),\, \alpha\,, \beta_i\}$ [note that in \eq(\ref{alpha}) the factor multiplying $V$ equals $\psi(0)$].
Solving \eq(\ref{betai}) for $\beta_i$ and replacing the result into \eq(\ref{alpha}) gives
\begin{equation}
\alpha\ug \frac{V\psi(0)}{\omega\meno\omega_0\meno\sum_i\tfrac{\mu_i^2}{\omega-\omega_i}}\label{alpha2}
\end{equation}
In the continuous limit, $\omega_i\!\rightarrow\!\omega'$, $\mu_i\!\rightarrow\!\mu(\omega')$ and $\sum_i\!\rightarrow\!\int{\rm d}\omega'\rho(\omega')$. Hence, using \eq (\ref{gee}) and $J(\omega)\ug \rho(\omega)\left[\mu(\omega)\right]^2$, \eq(\ref{alpha2}) becomes $\alpha\ug {V\psi(0)}\tilde\varepsilon(\omega)$ which once replaced into \eq(\ref{psip2}) finally yields
\begin{eqnarray}
\frac{d^2\psi}{dx^2}+k^2\,\psi(x)=2\frac{k}{\upsilon}\,W(\omega)\,\delta(x)\psi(x)\,\label{psi-eq}
\end{eqnarray}
with
\begin{eqnarray}
W(\omega)={V^2}\tilde\varepsilon(\omega)\label{gamma}\,,
\end{eqnarray}
where $\tilde\varepsilon(\omega)$ is the same function as in \eq(\ref{gee}) for $z\ug\omega$. 

The form of \eq (\ref{psi-eq}) is familiar in many contexts. In classical optics, an analogous equation describes an electromagnetic wave penetrating through a thin dielectric slab, \eg a {mirror} \cite{nota-mirror}. In elementary quantum mechanics, just the same equation is found for a particle of mass $k/\upsilon$ scattering from a potential barrier $W \delta(x)$.  
To make the language simpler, in what follows we refer to $W(\omega)$ as the {\it effective potential} (this is reminiscent of \rrefs\cite{ciccarello,sun}, where however $R$ was absent). Note that this is frequency-dependent, such dependance occurring through function $\varepsilon(\omega)$ which is associated with $J(\omega)$ [\cf\eq(\ref{gee})]. We have thus reduced our problem to the elementary calculation \cite{cohen-QM} of the reflection and transmission coefficients of a particle scattering from a pointlike potential barrier. These are given by \cite{note-rt}
\begin{eqnarray}
r=t-1=-\, \frac{i\,\frac{W}{\upsilon}}{1+i\,\frac{W}{\upsilon}}\label{rt}\,.
\end{eqnarray}

\section{Spectral density measurement}

It is important to stress that $W(\omega)$ is in general {\it complex}. Indeed, through the well-known Sokhotski--Plemelj theorem \cite{cohen, mertz} the improper integral appearing in \eq \eqref{gee}) for $z\ug\omega$ can be expressed as $\int \!d\omega'\,{J(\omega')}/({\omega\meno\omega'})\ug\mathcal{P}(\omega)\meno i\pi J(\omega)$, where $\mathcal{P}(\omega)$ stands for the integral's principal value. Thus $W(\omega)$ can be decomposed into its real and imaginary parts as
$W(\omega)=W_R(\omega)-i \,W_I(\omega)\label{gamma2}$ with
\begin{eqnarray}
W_R(\omega)&=&V^2\frac{ \omega-\omega_0-\mathcal P(\omega)}{[\pi 
   J(\omega)]^2+\left[\omega-\omega_0-\mathcal P(\omega)\right]^2}\,,\label{gammar}\\
W_I(\omega)&=&V^2\frac{\pi J(\omega)}{[\pi 
   J(\omega)]^2+\left[\omega-\omega_0-\mathcal P(\omega)\right]^2}\label{gammai}\,.
\end{eqnarray}
Note that $W_I(\omega)\!\ge\!0$. Using the aforementioned optical analogy, it is as if the photon impinges on an effective classical mirror, which besides being refractive is also {\it absorptive}. Scattering from complex potentials is a well-known tool \cite{schiff} arising as an effective description of inelastic scattering channels. Due to such channels, the sum of photon's reflection and transmission probabilities (reflectance and transmittance, respectively) is lower than one, namely $|r|^2\piu|t|^2\!<\!1$. From \eq(\ref{rt}), under the replacement $W\!=\!W_R\meno iW_I$, we indeed find
\begin{eqnarray}
1\meno|r|^2\meno|t|^2=\frac{2\,\frac{W_I}{\upsilon}}{\left(\frac{W_R}{\upsilon}\right)^2\piu\left(1\piu\frac{W_I}{\upsilon}\right)^2}\label{abs}\,,
\end{eqnarray}
which shows that $|r|^2\piu|t|^2\ug1$ if and only if $W_I\ug0$. The complexness of $W$ matches the physical expectation that -- due to the $R$ infiniteness -- the photon can be irreversibly absorbed by the $S$--$R$ system. 
A crucial point is that $W_I$ is strictly related to $J(\omega)$ [see \eq(\ref{gammai})]: the probability to lose the photon is non-zero whenever the SD at the photon's frequency $\omega$ is non-null, namely when $\omega$ matches a reservoir frequency (if any) that is coupled to $S$. We show next that, upon a perspective reversal, the last fact can be exploited for measuring the SD.

From \eqs(\ref{gammar}) and (\ref{gammai}) immediately follows that the SD can be expressed in terms of the effective potential $W(\omega)$ as
\begin{eqnarray}
J(\omega)=\frac{V^2}{\pi}\,\frac{W_I(\omega)}{|W(\omega)|^2}\,.\label{Jom2}
\end{eqnarray}
On the other hand, by taking the ratio of \eq (\ref{abs}) to the reflectance $|r|^2$ [\cf \eq(\ref{rt})], for any complex potential 
\begin{eqnarray}
\frac{W_I}{|W|^2}\ug\frac{1}{2\upsilon} \,\frac{1\meno|r|^2\meno |t|^2}{|r|^2}\label{gig2}\,.
\end{eqnarray}
Combined together, this and \eq(\ref{Jom2}) yield
\begin{eqnarray}
J(\omega)={\frac{V^2}{2\pi\upsilon}}\,\frac{1\meno|r(\omega)|^2\meno |t(\omega)|^2}{\,|r(\omega)|^2}\label{central}\,.
\end{eqnarray}
We thus find that the SD is a {\it universal simple function of the photon reflectance and transmittance}. A major immediate consequence of identity (\ref{central}) is that one can straightforwardly extract the SD profile from reflection and transmission spectra, which are normally easy to record via simple intensity measurements. 
Note that $V^2/\upsilon$ works as a sort of magnification knob: the larger $V^2/\upsilon$ the better the SD profile can be appreciated. This is reasonable since $V^2/\upsilon$ is the spontaneous emission rate of $S$ into the waveguide \cite{SF1}, i.e., it measures the effective $S$--$F$ coupling strength: a weakly (strongly) interacting photon is little (highly) sensitive to $S$--$R$. Note that no approximations on the $S$-$R$ coupling strength has been made to derive \eq(\ref{central}). Hence, it is not limited to the weak-coupling regime. Indeed, in the illustrative process discussed after \eq(\ref{FGR}) -- where a FGR-based measurement fails -- \eq(\ref {central}) correctly predicts $J(\omega)\ug0$ for $\omega$ within a band gap of $R$ (no matter how close to the edge): at such photon frequency $\Gamma(\omega)$ is real, thereby $1\meno|r(\omega)|^2\meno |t(\omega)|^2\ug0$ [\cf \eqs(\ref{gammai}) and \eqref{abs}].

A formally trivial, yet physically noteworthy, consequence of \eq(\ref{central}) is that for a {\it flat} SD the combination of reflectance and transmittance 
\begin{equation}
f(\omega)\ug\frac{1\meno|r(\omega)|^2\meno|t(\omega)|^2}{|r(\omega)|^2}\label{fomega}
\end{equation} 
is constant in frequency. Given that a constant SD yields a Markovian dynamics described by the KLME,  the flatness of function $f(\omega)$ can be used as a test to assess whether the KLME is effective.

\section{Test of \eq(\ref{central}) based on experiments}

To provide a check of \eq(\ref{central}) based on real experiments, let us consider microwave photon scattering in a 1D open transmission line from a superconducting qubit (artificial atom), which has been the focus of \rrefs \cite{tsai,delsing-NJP}. In these experiments, it was found that the measured field reflection and transmission coefficients are well described by
\begin{eqnarray}
r(\omega)\ug t(\omega)\meno1\ug-\frac{\Gamma_{eg}}{2\gamma}\,\frac{1\meno i\delta\omega/\gamma}{1\piu(\delta\omega/\gamma)^2+\Omega^2/[(\Gamma_{eg}\piu\Gamma_l)\gamma]}\label{rt-tsai}\,,\,\,\,\,\label{rtmw}
\end{eqnarray}
where $\delta\omega\ug\omega\meno\omega_0$, $\Gamma_{eg}$ is the atom's relaxation rate into the waveguide modes while $\gamma\ug\Gamma_{eg}/2\piu\Gamma_{\phi,l}$ is the total atom's decoherence rate. Here, importantly, $\Gamma_{\phi,l}\ug\Gamma_l/2\piu\Gamma_\phi$ depends on the atom's coupling to reservoirs external to the waveguide modes: $\Gamma_l$ is the rate of intrinsic losses while $\Gamma_\phi$ is the pure dephasing rate. $\Omega$ is the Rabi frequency proportional to the input field power. Computing $f(\omega)$ as defined in \eq(\ref{fomega}), we find
\begin{eqnarray}
f(\omega)\ug\frac{2 (\Gamma_l\piu2 \Gamma_{\phi})}{\Gamma_{eg}}\piu
\frac{4 (\Gamma_{eg}\piu\Gamma_l\piu2
   \Gamma_{\phi})^2\,\Omega^2}{\Gamma_{eg}
   (\Gamma_{eg}\piu\Gamma_l)
   \left[(\Gamma_{eg}\piu\Gamma_l\piu2 \Gamma_{\phi})^2\piu4
   \delta\omega^2\right]} .\nonumber
\end{eqnarray}
For a single-photon beam, $\Omega$ is negligible \cite{tsai,delsing-NJP} and $f(\omega)$ becomes independent of $\omega$ (the second term vanishes). In our framework, this corresponds to a flat SD yielding that the atom's open dynamics is describable through the KLME. Significantly, this is {consistent} with coefficients (\ref{rtmw}) since these can be worked out through a simple semiclassical model \cite{tsai} based on a KLME for the atom's density matrix (the input field being treated as a classical drive). Note that for the present setting, besides the dissipative reservoir associated with rate $\Gamma_l$, the atom is subject also to purely dephasing noise (corresponding to $\Gamma_\phi$). While samples with negligible $\Gamma_\phi$ are within reach \cite{delsing-NJP}, being thus fully compatible with Hamiltonian (\ref{SR}), it is remarkable that $f(\omega)$ for $\Omega\!\simeq\!0$ is flat even for $\Gamma_\phi\!\neq\!0$ (suggesting that the scheme might be generalizable to some extent). 

To test \eq(\ref{central}) in the case of a structured SD, let us consider the experiment in \rref\cite{wallraf}. This differs from the previous one for the fact that the OQS coupled to the 1D line (i.e., system $S$) is a high-finesse resonator $C$. Also, the resonator is coupled to a lossy Cooper pair box (CPB) via a Jaynes-Cummings (JC) interaction \cite{barnett}. Here, $R$ is jointly embodied by the CPB {\it and} its own reservoir. Note that in such single-photon process, both $C$ and the CPB behave as effective qubits. In \rref\cite{SF-PRA}, it was shown that the scattering coefficients in such experiment are reasonably given by
\begin{eqnarray}
r(\omega)\ug t(\omega)-1\ug-i\frac{V^2}{\upsilon}\frac{ \omega\meno \omega_1\piu i \Gamma_1}{
   (\omega\meno\omega_{1}\piu i \Gamma_1)
   \left(\omega\meno\omega_0\piu\frac{i
   V^2}{\upsilon}\right)\meno g^2}\,,\label{rt-wall}
   \end{eqnarray}
where $\omega_0$ ($\omega_{1}$) is the frequency of $C$ (CPB), $\Gamma_1$ is the CPB dissipation rate and $g$ is the $C$-CPB coupling rate. In such case, \eq(\ref{central}) yields the Lorentzian SD
\begin{eqnarray}
J(\omega)\ug\frac{g^2}{\pi}\frac{\Gamma_1}{\Gamma_1^2+(\omega-\omega_{1})^2}\,,\label{SD-W}
\end{eqnarray}
which is indeed a signature of a damped JC dynamics \cite{petruccione,lambro}  (observing a JC coupling is the main focus of \rref\cite{wallraf}).

It should be noted that \eq(\ref{psi-eq}), hence \eq(\ref{central}), relies on the assumption (which routinely occurs in waveguide QED and beyond) that only a relatively narrow photonic bandwidth is involved \cite{SF-PRA}. This could be no more valid when the dressing of $S$ by $R$ is very strong. Yet, this does not prevent \eq\eqref{central} from holding for significantly strong $S$-$R$ couplings. In this respect, we note that for the SD \eq\eqref{SD-W} various non-Markovianity measures [32] are non-zero in the resonant case $\omega_0\ug\omega_1$ if and only if 4$g^2/\Gamma_1^2\! >\! 1$. In the setup of \rref[19] discussed above, $g/(2\pi)\!\simeq\!5.8\,$MHz while $\Gamma_1/(2\pi)\!\simeq\!0.7\,$MHz so that  one can estimate $4g^2/\Gamma_1^2\!\simeq\!275$, which by far exceeds the above threshold.

\section{Conclusions}

In this paper, we have shown a method for measuring the SD of a reservoir in dissipative contact with a small OQS. This is achieved by coupling the OQS to a 1D photonic waveguide and sending through this photons which undergo scattering from the OQS. The SD has been shown to be a simple universal function of photon reflectance and transmittance. As such, it can be easily extracted from reflection and transmission spectra. The result  does not rely on the weak-coupling approximation. The SD measurement is ``static" since the dynamical evolution of the OQS in contact with $R$ needs not be switched on or monitored. Such dynamics is fully reconstructable via the 1D photon scattering. The scheme diagnostic power has been tested on the basis of two real waveguide-QED experiments, including one exhibiting a structured SD \cite{nota-test}.

A remarkable point is that while -- expectably -- the scattering coefficients depend on the SD in quite a complicated way [\cf\eqs(\ref{gee}), (\ref{gamma})-(\ref{rt})] the SD is instead quite a simple function of them. We envisage that this property, suggesting the perspective reversal at the heart of the method, has the potential to inspire novel approaches to the SD measurement problem in more general situations such as finite-temperature and/or purely dephasing noise.

Fruitful discussions with T. Tufarelli, S. Lorenzo, M. Paternostro, G. M. Palma, R. Lo Franco and L. Chirolli are gratefully acknowledged.


\newpage





\begin {thebibliography}{99}
\bibitem{haroche} S. Haroche, Rev. Mod. Phys. {\bf 85}, 1083 (2013); D. J. Wineland, Rev. Mod. Phys. {\bf 85}, 1103 (2013).
\bibitem{petruccione} H. P. Breuer and F. Petruccione, {\it The Theory of Open
Quantum Systems} (Oxford, Oxford University Press, 2002).
\bibitem{cohen} C. Cohen-Tannoudji, J. Dupont-Roc, and G. Grynberg, {\it AtomÐPhoton Interactions} (New York, Wiley, 1992).
{\bibitem{huelga} A. Rivas and S.F. Huelga, {\it Open Quantum Systems. An Introduction} (Springer, Heidelberg, 2011).}
\bibitem{breuer} H.-P. Breuer, J. Phys. B: At. Mol. Opt. Phys. {\bf 45}, 154001 (2012).
\bibitem{rivas} A. Rivas, S.F. Huelga, and M. B. Plenio, Rep. Prog. Phys. {\bf 77}, 094001 (2014).
\bibitem{lambro} P Lambropoulos, G. M. Nikolopoulos, T. R. Nielsen, and S. Bay, Rep. Prog. Phys. {\bf 63}, 455 (2000).
\bibitem{shabani} A. Shabani, M. Mohseni, H. Rabitz, and S. Lloyd, Phys. Rev. E {\bf 86}, 011915 (2012).
\bibitem{nc} M. A. Nielsen and I. L. Chuang,  \textit{Quantum Computation and Quantum Information} (Cambridge University Press, Cambridge, U. K., 2000).
\bibitem{aspuru} T. Markovich, S. M. Blau, J. Parkhill, C. Kreisbeck, J. N. Sanders, X. Andrade, and A. Aspuru-Guzik, arXiv:1307.4407.
\bibitem{dissi} A. A. Clerk, M. H. Devoret, S. M. Girvin, F. Marquardt, and R. J. Schoelkopf, Rev. Mod. Phys. {\bf 82}, 1155 (2010); R. J. Schoelkopf, A. A. Clerk, S. M. Girvin, K.W. Lehnert, and M. H. Devoret, arXiv: cond-mat/0210247; 
\bibitem{dephase} See \eg G. A. Alvarez and D. Suter, Phys. Rev. Lett. {\bf 107}, 230501 (2011); J. Bylander {\it et al.}, Nat. Phys. {\bf 7}, 565 (2011); K. C. Young and K. B. Whaley, Phys. Rev. A {\bf 86}, 012314 (2002); T. Fink and H. Bluhm, Phys. Rev. Lett.  {\bf 110}, 010403 (2013).
\bibitem{SF1}
J.-T. Shen and S. Fan, Opt. Lett. {\bf 30}, 2001 (2005); Phys. Rev.
Lett. {\bf 95}, 213001 (2005).
\bibitem{SF-PRA} J. T . Shen and S. Fan, \pra {\bf 79}, 023837 (2009).
\bibitem{switch} L. Zhou, Z. R. Gong, Y. X. Liu, C. P. Sun, and F. Nori, Phys. Rev. Lett. {\bf 101}, 100501 (2008).
\bibitem{lukin} D. E. Chang, A. S. Sorensen, E. A. Demler, and M. D. Lukin, Nat. Phys. {\bf 3}, 807 (2007).
\bibitem{sore} D. Witthaut and A. S. Sorensen, New J. Phys. {\bf 12}, 043052 (2010).
\bibitem{review} P. Lodahl, S. Mahmoodian, and S. Stobbe, arXiv:1312.1079 (2013).
\bibitem{wallraf} A. Wallraff {\it et al.}, Nature (London) {\bf 431}, 162 (2004).
\bibitem{tsai} O. V. Astafiev, A. M. Zagoskin, A. A. Abdumalikov, Jr., Y. A. Pashkin, T. Yamamoto, K. Inomata, Y. Nakamura, and J. S. Tsai, Science {\bf 327}, 840 (2010).
\bibitem{delsing-NJP} I.-C. Hoi, C. M. Wilson, G. Johansson, J. Lindkvist, B. Peropadre, T. Palomaki, and P. Delsing, New J. Phys. {\bf 15} 025011 (2013).
\bibitem{claudon} T. Lund-Hansen {\it et al.}, Phys. Rev. Lett. {\bf 101}, 113903 (2008); J. Claudon {\it et al.}, Nat. Photonics {\bf 4} 174 (2010); T. M. Babinec {\it et al.}, Nat.
Nanotechnol. {\bf  5}, 195 (2010); A. Laucht {\it et al.}, Phys. Rev. X {\bf 2}, 011014 (2012); A. Akimov A. {\it et al.}, Nature {\bf 450}, 402 (2007); A. Huck, S. Kumar, A. Shakoor and U. L. Andersen, Phys. Rev. Lett., {\bf 106}, 096801 (2011);  M. Arcari {\it et al.}, Phys. Rev. Lett. {\bf 113}, 093603, (2014).
\bibitem{ciccarello} F. Ciccarello, D. E. Browne, L. C. Kwek, H. Schomerus, M. Zarcone, and S. Bose, Phys. Rev. A {\bf 85}, 050305(R) (2012).
\bibitem{zheng} H. Zheng, D. J. Gauthier, and H. U. Baranger, Phys. Rev. Lett. {\bf 111}, 090502 (2013).
\bibitem{nota-FGR} To obtain \eq(\ref{FGR}), one approximates the integral appearing in \eq \eqref{gee} by replacing $z\!\rightarrow\!\omega_0$ (pole approximation \cite{lambro}). This yields  $\int \!d\omega'{J(\omega')}/({z\meno\omega'})\!\simeq\!\mathcal{P}(\omega_0)\meno i\pi J(\omega_0)$ with $\mathcal{P}(\omega)$ the integral principal value. Through inverse LT of \eq(\ref{gee}), we thus find $|\varepsilon(t)|^2\!\simeq\!e^{-2\pi J(\omega_0)t}$.
\bibitem{kurizki} A. G. Kofman, G. Kurizki, and B. Sherman, J. Mod. Opt. {\bf 41}, 353 (1994).
\bibitem{sun} Z. R. Gong, H. Ian, L. Zhou, and C. P. Sun, Phys. Rev. A {\bf 78}, 053806 (2008)
\bibitem{nota-mirror} The same equation can indeed be found for a single photon if the last integral in (\ref{H}) were replaced with $\!\int \!{\rm d}x\,W\delta(x)\hat c^\dag(x)\hat c(x)$ (and $\hat H_{S\!R}$ set to zero). 
\bibitem{cohen-QM} C. Cohen-Tannoudji, B. Diu, and F. Laloe, {\it Quantum Mechanics} (Paris, Wiley Interscience, 1977).
\bibitem{note-rt} Coefficients (\ref{rt}) are found by imposing that $\psi(x)$ be continuous  at $x\ug0$ and $\partial_x\psi(0^+)\meno\partial_x \psi(0^-)\ug 2 k/\upsilon\,W\psi(0)$ (the latter constraint is obtained upon integration of \eq(\ref{psi-eq}) over an infinitesimal interval across $x\ug0$ \cite{cohen-QM}). 
\bibitem{mertz} E. Merzbacher, {\it Quantum Mechanics} (New york, John Wiley \& Sons, 1970).
\bibitem{schiff} See \eg L. I. Schiff, {\it Quantum Mechanics} (New York,  McGraw Hill, 1968)
\bibitem{barnett} S. M. Barnett and P. M. Radmore, {\it Methods in Theoretical Quantum Optics} (Oxford, Clarendon Press, 1997).
\bibitem{measures} H.-P. Breuer, E.-M. Laine, and J. Piilo, Phys. Rev. Lett. {\bf 103}, 210401 (2009); A. Rivas, S. F. Huelga and M. Plenio, Phys. Rev. Lett. {\bf 105}, 050403 (2010); S. Lorenzo, F. Plastina and M. Paternostro, Phys. Rev. A {\bf 88}, 020102(R) (2013).
\bibitem{nota-test} Formally, a check of \eq(\ref{central}) in the addressed instances of a flat and Lorentzian SD [\cf\eqs(\ref{rt-tsai}) and (\ref{rt-wall}), respectively] might appear redundant. Yet, one has to consider 
that \eqs(\ref{rt-tsai}) and (\ref{rt-wall}) reflect the outcomes of {\it real} experiments, which dresses the performed tests of physical significance.
\end {thebibliography}
\end{document}